\documentclass[twocolumn]{aastex631}

\usepackage{tabularx}
\usepackage{amsmath}
 
\usepackage{graphicx}
\usepackage{booktabs}
\usepackage{soul}

\newcommand\p{\partial}

\shorttitle{Nonthermal distributions}
\shortauthors{Afify et al.}

\begin{document}
%\linenumbers

\title{Impact of nonthermal electron distributions on the triggering of the ion-ion acoustic instability near the Sun: Kinetic simulations }
\correspondingauthor{Mahmoud Saad Afify}
\email{Mahmoud.Ibrahim@ruhr-uni-bochum.de\\Mahmoud.Afify@fsc.bu.edu.eg}

\author[0000-0002-2088-0354]{Mahmoud Saad Afify}
\affiliation{Theoretische Physik I, Ruhr-Universität Bochum, Bochum, Germany}
\affiliation{Department of Physics, Faculty of Science, Benha University, Benha 13518, Egypt}
\author[0000-0002-6189-9158]{Jürgen Dreher}
\author[0000-0002-4136-4244]{Stuart O'Neill}
\author[0000-0002-5782-0013]{Maria Elena Innocenti}
\affiliation{Theoretische Physik I, Ruhr-Universität Bochum, Bochum, Germany}

\begin{abstract}
\par\noindent \textit{Context.} In a previous paper \citep{Afify2024}, we have investigated the stability threshold of the ion-ion acoustic instability (IIAI) in parameter regimes compatible with recent Parker Solar Probe (PSP,~\citep{Fox2016}) observations, in the presence of a Maxwellian electron distribution. We found that observed parameters are close to the instability threshold, but IIAI requires a higher electron temperature than observed.   \\
\textit{Aims.} As electron distributions in the solar wind present clear non-Maxwellian features, we investigate here if deviations from the Maxwellian distribution could explain the observed IIAI. We address specifically the kappa ($\kappa$) and core-strahl distributions for the electrons.\\
\textit{Methods.} We perform analytical studies and kinetic simulations using a Vlasov-Poisson code in a parameter regime relevant to PSP observations. The simulated growth rates are validated against kinetic theory.\\
\textit{Results.} We show that the IIAI threshold changes in the presence of $\kappa$ or core-strahl electron distributions, but not significantly. In the latter case, the expression of an effective temperature for an equivalent Maxwellian electron distribution given in~\citet{jones1975propagation} is confirmed by simulations. Such an effective temperature could simplify stability assessment of future observations.   

\end{abstract}

\section{Introduction} \label{sec:intro}
Observations of plasma in the solar wind and planetary magnetospheres over the past decades have made it clear that non-Maxwellian distribution functions are ubiquitous \citep{Feldman1973, Feldman1974, Feldman1975, maksimovic1997ulysses, Marsch1982, Marsch1987, Neugebauer1996, marsch2006kinetic, Marsch2018, Klein2018, Durovcova2019}. 
These nonthermal structures actively shape the large-scale dynamics of the solar wind by driving microinstabilities \citep{vstverak2009radial, Matteini2013} which in turn constrain large-scale solar wind parameters, influencing wave spectra \citep{verscharen2011apparent, yoon2024non}, and mediating particle transport \citep{marsch2006kinetic, reames2021solar}.

In the solar wind, in-situ measurements reveal that proton velocity distribution functions (VDFs) frequently deviate from Maxwellian equilibrium distributions. A common non-equilibrium feature is the presence of a field-aligned beam, which is a secondary proton population streaming faster than the core proton component along the magnetic field direction \citep{Feldman1974, Marsch1982, alterman2018comparison, Verniero2020, Verniero2022}. Moreover, proton VDFs often exhibit temperature anisotropies with respect to the local magnetic field \citep{marsch1981pronounced, marsch2004temperature, Hellinger2006solar, bale2009magnetic, maruca2012instability}. All these features are generally more pronounced in fast solar wind compared to slow wind \citep{Verscharen2019}.  The presence of a (possibly anisotropic) ion beam plays a crucial role in driving a variety of ion-scale instabilities \citep{GaryBook, gary1984electromagnetic, gary1985electromagnetic, daughton1998electromagnetic, Verscharen2013, verscharen2013dispersion, gary2016ion}. 
Extensive investigations of ion beam instabilities have been conducted using both hybrid simulations \citep{daughton1999electromagnetic, gary2000electromagnetic, wang2003generation, lu2006hybrid, ofman2022modeling, ofman2023observations, ofman2025evidence} and fully kinetic particle-in-cell (PIC) approaches~ \citep{riquelme2015particle, che2023particle, Pezzini2024}. However, these studies have primarily focused on ion kinetic physics while often neglecting the effects of the non-Maxwellian electron VDFs commonly observed in the solar wind. The works that have examined the effect of non-Maxwellian electron velocity distribution functions on ion-scale instabilities have usually focused on anisotropic, Maxwellian electron distributions~\citep{ahmadi2016effects, micera2020particleFirehose, walters2024electron}. 
%The solar wind further displays anisothermal behavior, where different plasma species maintain unequal temperatures \citep{feldman1974solar, cohen1996kinetic, von2006kinetic}. 
%Notably, electrons are typically cooler than protons in fast solar wind but hotter than protons in slow solar wind \citep{ newbury1998electron}. 
%Electrons, despite their low mass and negligible contribution to the solar wind’s momentum flux, play a crucial role in maintaining quasi-neutrality over spatial scales much larger than the electron Debye length and on timescales longer than the inverse electron plasma frequency \citep{salem2003electron}. Their high mobility and thermal velocity, significantly greater than that of protons, result in subsonic behavior, with drift velocities remaining well below their thermal speed.

Solar wind electron velocity distributions also exhibit distinct nonthermal characteristics. They typically comprise a thermal core component and a field-aligned, anti-Sunward propagating beam, the strahl \citep{Feldman1973, feldman1974solar, Feldman1975, pilipp1987characteristics, lin1998wind, maksimovic2005radial, vstverak2009radial, micera2020particle, Micera2021, micera2025quasi}. The core population, representing approximately 95 \% of the total electron density, dominates the distribution. The strahl is composed of high-energy electrons that escaped solar gravity along open magnetic field lines, and got focused about the field direction by conservation of the first adiabatic invariant in a magnetic field of decreasing magnitude~\citep{meyer2007basics}. %They manifest as a pronounced, magnetic-field-aligned shoulder in the distribution function at small pitch angles in the antisunward direction \citep{ fitzenreiter1998observations}.
%The transition energy between the core and strahl populations, along with their relative densities, demonstrates a clear dependence on heliocentric distance and shows systematic correlations with solar wind speed and temperature \citep{ maksimovic2005radial, vstverak2009radial, graham2017evolution, abraham2022radial}. These populations typically exhibit non-zero relative drifts along the magnetic field direction. To maintain global quasi-neutrality as required by Poisson's equation, the field-aligned drifts must balance such that the net electron charge flux equals the ion charge flux. This condition generally results in a sunward drift of the core population that compensates for the antisunward-directed strahl.
%The relative drifts between electron populations, particularly those involving suprathermal components, constitute the primary mechanism for electron heat flux transport. Near $r \approx 1$ AU, a suprathermal halo population emerges \citep{Feldman1975, pilipp1987characteristics, lie1997kinetic, maksimovic1997ulysses}. 
A third, thin, hot and often $\kappa$-distributed population, the halo, originates by scattering of the strahl population by instabilities in the whistler family, as demonstrated by the anti-correlation between halo and strahl density~\citep{maksimovic2005radial}, direct in-situ PSP observations~\citep{cattell2021parker}, and Particle In Cell simulations~\citep{micera2020particle, Micera2021}.

%To quantitatively model these non-equilibrium characteristics in solar wind velocity distribution functions (VDFs), empirically derived distributions, including the bi-Maxwellian \citep{marsch2006kinetic} and kappa \citep{summers1991modified, vasyliunas1968survey} formulations, are widely adopted.
%The Kappa distribution, in particular, has been widely used to assess instability thresholds as a function of proton temperature.
The $\kappa$-distributions are a type of nonthermal distribution characterized by a Maxwellian-like core and power-law tails that represent an enhanced population of high-velocity particles \citep{summers1991modified, vasyliunas1968survey, maksimovic1997kinetic}. The suprathermal characteristics of the distribution increase as the parameter $\kappa$ decreases, with the distribution becoming nonphysical at the critical value $\kappa = 3/2$ where the mean kinetic energy becomes infinite, i.e. the superthermal distribution becomes non-normalizable in terms of finite energy \citep{pierrard2010kappa}. At $\kappa=\infty$, a Maxwellian distribution is recovered. Due to its convenience in modelling both the core and superthermal populations of the solar wind, as well as the advantage of recovering a Maxwellian distribution at the high $\kappa$ limit, the distribution has been extensively used in kinetic models of the solar wind \citep{maksimovic1997ulysses}.
 
Parker Solar Probe (PSP) has provided invaluable insights into ion-scale instabilities in the solar wind. In fact, PSP observations have shown the ubiquity of ion-scale wave activity, including a number of ``ion storms" related to the presence of ion beams and anisotropies \citep{Verniero2020, Verniero2022}.   
%Through its observations, the Parker Solar Probe (PSP) \citep{Fox2016} has provided valuable insights into the properties of ion-acoustic waves (IAWs) over a range of heliocentric distances
While most ion instabilities are electromagnetic in nature, the electrostatic Ion-Ion Acoustic Instability, IIAI, has been observed and characterized over a range of heliocentric distances~\citep{Mozer2020, mozer2021triggered, Mozer2023, Mozer2025}. The IIAI is driven by a proton core and beam drifting with respect to each other, in the presence of high temperature electrons to minimize Landau damping \citep{Gary1987, Silin1992, Afify2024}  
%Several mechanisms may account for the observed IAWs \citep{GaryBook, galeev1979nonlinear}:(i) Ion-acoustic instability (IIAI) is a counter-streaming kinetic instability and represents a compelling mechanism for explaining observed IAWs. The excitation threshold for IIAI requires the relative drift between charged particles to exceed the ion-acoustic speed. This instability exhibits a strong dependence on the electron-to-ion temperature ratio ($T_e/T_i \gg 1$), as lower ratios lead to significant wave damping \citep{Gary1987, Gurnett1991}.(ii) At larger heliocentric distances, current-driven instabilities may trigger IIAI when the high temperature ratio condition is satisfied \citep{Lemons1979}. In such cases, Landau resonance, driven by a small population of resonant electrons, can generate IAWs. (iii) Additional excitation mechanisms include nonlinear wave–wave interactions, such as Langmuir wave decay and whistler-mode turbulence \citep{Cairns1992, Saito2017}.(iv) Recent observations particularly support the ion–ion-acoustic instability mechanism \citep{Gary1987, Silin1992, Afify2024} as an explanation for IAW propagation in the solar wind. This interpretation is reinforced by in-situ measurements showing plateau formation in ion velocity distribution functions during wave activity, indicative of ion beam relaxation processes \citep{Mozer2021a}.
In our previous work,~\citet{Afify2024} we have investigated IIAI onset in parameter regimes comparable with the observations in~\citet{mozer2021triggered}. Through combined theoretical and simulation analysis, we have demonstrated that the solar wind parameters reported in~\citet{mozer2021triggered} are \textit{in the vicinity of the IIAI threshold}, but not in the unstable regime yet. In particular, we have succeeded in reproducing frequency, wavenumber and magnitude of the high frequency IIAI observed there, but only after (slightly) modifying key parameters such as the electro-to-proton-core temperature ratio, the ratio between the parallel beam-to-proton-core temperature, and the relative drift between the core and beam protons. All our investigations have been conducted in the presence of a Maxwellian electron population, which, as already mentioned, does not reflect the observed electron distribution in the solar wind.
Our aim now is to investigate if non-Maxwellian electron VDFs can promote the onset of the ion-ion acoustic instability in parameter regimes compatible with those observed in \citet{mozer2021triggered}. 
%As discussed in \citep{Afify2024}, these parameters are next to instability threshold, but not in the unstable regime. 
To do so, we start from one of the cases analyzed in \citet{Afify2024}, where the temperature ratio between the electron ($e$) and proton core ($c$) population had been increased from the observed $T_e/T_c \sim 7$ to $T_e/T_c= 10$. Similarly, the temperature ratio between the proton beam ($b$) and core component has been decreased from the observed $T_b/T_c=2.7$ to  $T_b/T_c=1$.
We consider two types of electron distributions. In Sec. \ref{sec:kappa}, we start with $\kappa$-distributed electrons, often used to approximate the observed core plus suprathermal electron distribution. We examine the effect of the $\kappa$ parameter on the instability threshold. Then, we move to a core-strahl distribution, both Maxwellian in Sec.~\ref{sec:core-strahl}. We investigate the IIAI threshold variation as a function of the electron strahl's density and temperature, and of the relative drift speed between the proton core and beam. Discussion and conclusions are presented in Sec.~\ref{sec:concl}. 

\section{Kappa-distributed electrons}
\label{sec:kappa}
\begin{figure*}  [h!]
\begin{minipage}[b]{0.48\textwidth}
    \includegraphics[width=\textwidth]{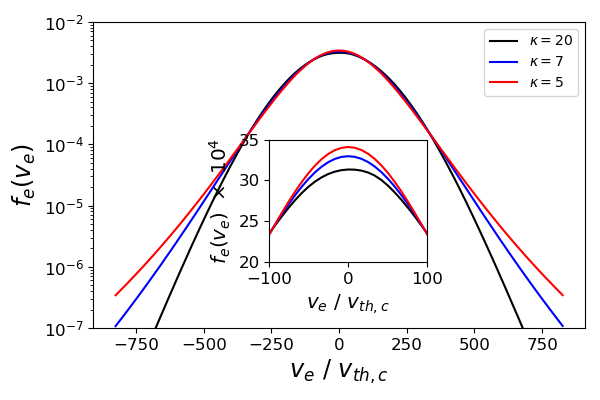}
 \end{minipage}
\begin{minipage}[b]{0.48\textwidth}
    \includegraphics[width=\textwidth]{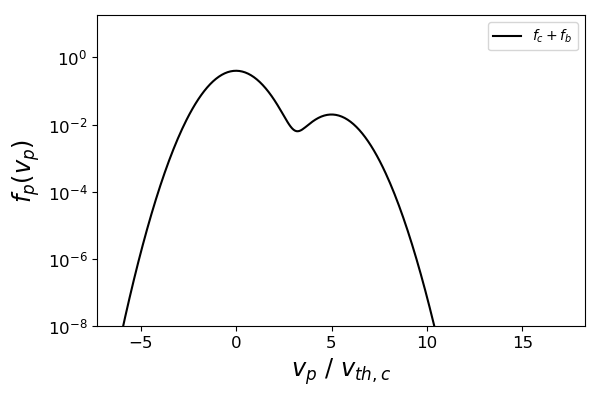}
 \end{minipage}

\caption{Electron and proton distribution functions. Left panel: $\kappa$ electron distributions with $T_e/T_c=10$, $\kappa=20$ (black curve), $\kappa=7$ (blue curve), and $\kappa=5$ (red curve). Right panel: total proton distribution consisting of two Maxwellians (core and beam), with $V_{d}/ v_{th,c}= 5$, $T_b/ T_c= 1.0$, and $n_b/n_c=0.05$.} 
\label{VDF_kappa}
\end{figure*}

\begin{figure*}  [h!]
\begin{minipage}[b]{0.48\textwidth}
   \includegraphics[width=\textwidth]{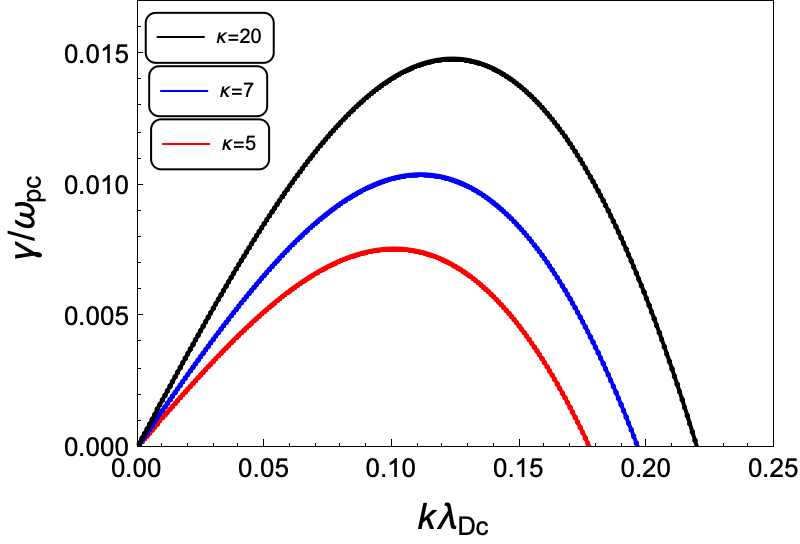}
 \end{minipage}
\begin{minipage}[b]{0.48\textwidth}
    \includegraphics[width=\textwidth]{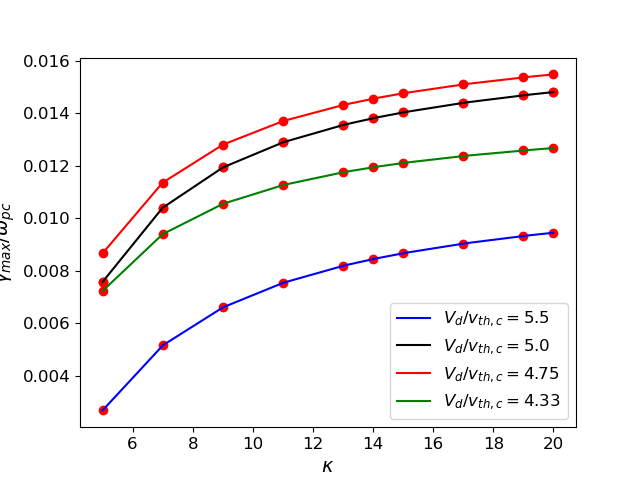}
 \end{minipage}
\caption{Dispersion and growth characteristics of the IIAI. (Left panel): Dispersion relation for the IIAI with $\kappa$-distributed electrons and $V_d/v_{th,c}=5$, $n_b/n_c=0.05$, $T_e/ T_c=10$, $T_b/T_c=1$. The black, blue, and red lines are $\kappa= 20, 7,$ and $5$ respectively.  (Right panel): Normalized maximum growth rates, $\gamma_{max}/ \omega_{pc}$, as a function of the $\kappa$ for different proton core-beam drift speeds $V_d/v_{th,c}=4.33$ (green line), $4.75$ (red line), $5.0$ (black line), and $5.5$ (blue line). Remaining parameters ($n_b/n_c$, $T_e/ T_c$, and $T_b/T_c$) are the same as in panel a. }
%\textcolor{red}{in the legend of panel b, use $\Delta v_d/v_{th,c}$ not v. } }
\label{DR_kappa}
\end{figure*}

We first examine the impact of $\kappa$-distributed, as opposed to Maxwellian-distributed electrons on the IIAI. We choose the 1D standard $\kappa$ distribution given by \citet{vasyliunas1968survey, summers1991modified, abdul2014method}
\begin{equation}
f\left(v\right)=\left(\pi \kappa \theta^2\right)^{-1 / 2} \frac{\Gamma(\kappa)}{\Gamma(\kappa-1 / 2)}\left(1+\frac{v^2}{\kappa \theta^2}\right)^{-\kappa},
\end{equation}
where $\theta^2=2[(\kappa-3 / 2) / \kappa] v_{t h,e}^2$ is the generalised thermal velocity, which is a function of the $\kappa$ index. Here, $v_{th,e} = \sqrt{\frac{T_e}{m_e}}$ is the electron thermal velocity, with $T_e$ expressed in energy units, representing the average kinetic energy per particle. %This definition of temperature aligns with that of the Maxwellian distribution, making comparison straightforward \citep{lazar2015destabilizing}.
Figure \ref{VDF_kappa} shows the velocity distributions of electrons and protons in our model of the non-thermal solar wind plasma. The left panel displays $\kappa$-distributed electrons for three different spectral indices $\kappa = 20$, $\kappa = 7$, and $\kappa = 5$, with electron to proton core temperature $T_e/T_c=10$. As $\kappa$ decreases, the distribution develops a more pronounced high-energy tail, characteristic of suprathermal populations. At the same time, the peak electron density at $v_e/ v_{th,c} \sim 0$ increases (see inset). The right panel shows the combined distribution of protons, consisting of a Maxwellian-distributed core and beam component for an example case. Motivated by PSP observations reported in \citet{mozer2021triggered}, we choose a proton beam-core drift speed  of $V_d  = 5\; v_{th,c}$, with $v_{th,c}$ the thermal speed of the proton core, equal beam and core temperatures ($T_b = T_c $), a dilute proton beam ($n_b / n_c = 0.05$), with $n_j$ the density of the population $j$ and $b$ the proton beam. These parameters are in the range of PSP observations close to the Sun \citep{mozer2021triggered}.
We first calculate the dispersion relation of the IIAI, given by Eq. (3), as in \citet{Afify2024}, but now with $\kappa$-distributed electrons. The plasma dispersion function \citep{fried1961plasma} is evaluated numerically using the Faddeeva function as implemented in SciPy to find its roots \citep{virtanen2020scipy}.
Figure \ref{DR_kappa}, panel a, shows the dispersion relation for $\kappa=20,7,5$ in black, blue, and red, respectively. $\gamma$ is the growth rate, normalized to the core proton plasma frequency $\omega_{pc}$. The wavenumber $k$ is normalized to the core proton Debye length $\lambda_{Dc}$. The parameters ($V_d/ v_{th,c}$, $n_b/ n_c$, $T_e/ T_c$, $T_b/T_c$) are the same as mentioned above.
We described in~\citet{Afify2024} how the maximum IIAI growth rate first increases and then decreases with increasing drift velocity in the case of Maxwellian electron distribution. This feature is also found with $\kappa$-distributed electrons, as shown in Fig.~\ref{DR_kappa}, panel b.
%The variation of the maximum growth rate for $5 \leq \kappa \leq 20$ is depicted in Figure \ref{DR_kappa}, panel b for four different values of the $\Delta v_d/ v_{th,c}$ parameter. %There, the red, black, green, and blue lines are $\Delta v_d/ v_{th,c}=4.75, 5, 4.33,$ and $5.5$, respectively. %Again,  $n_b/ n_c=0.05$, $T_e/ T_c=10$, $T_b/T_c=1$. 
%The horizontal lines are the growth rates in the presence of a Maxwellian electron population, as already examined in \citep{Afify2024}, Figure 1 (dashed line). 
%Notice that the maximum growth rate first increases and then decreases with increasing drift velocity due to the variation of the instability threshold with drift velocity, as already explained in \citet{Afify2024}.
We see from both panels of Fig.~\ref{DR_kappa} that for $\kappa$-distributed electrons, smaller $\kappa$ values reduce the  growth rate of the instability. 
%A natural explanation could be enhanced Landau damping at smaller kappa values, due to higher electron density at the resonant velocity $v_{res}= \omega_{r}/ k << v_{th,e}$, as seen in Fig.~\ref{VDF_kappa}.
%\st{Figure \ref{kappaArguments} presents the influence of superthermal electrons on the excitation of IIAI. The results demonstrate that beam protons (upper panel) with lower $|\zeta_b| > 1$ and electrons (lower panel) with lower $|\zeta_e| < 1$ exhibit stronger resonance at higher $\kappa$ values. In contrast, core protons (middle panel) show weaker resonance at higher $\kappa$ values, particularly for larger $|\zeta_c| > 1$. Consequently, the presence of $\kappa$-distributed electrons modifies the dispersion properties, suppressing the resonance of IAWs with core protons while enhancing electron resonance, thereby facilitating the onset of IIAI.}

\begin{figure*}  %[h!]
\centering
\includegraphics[scale=0.58]{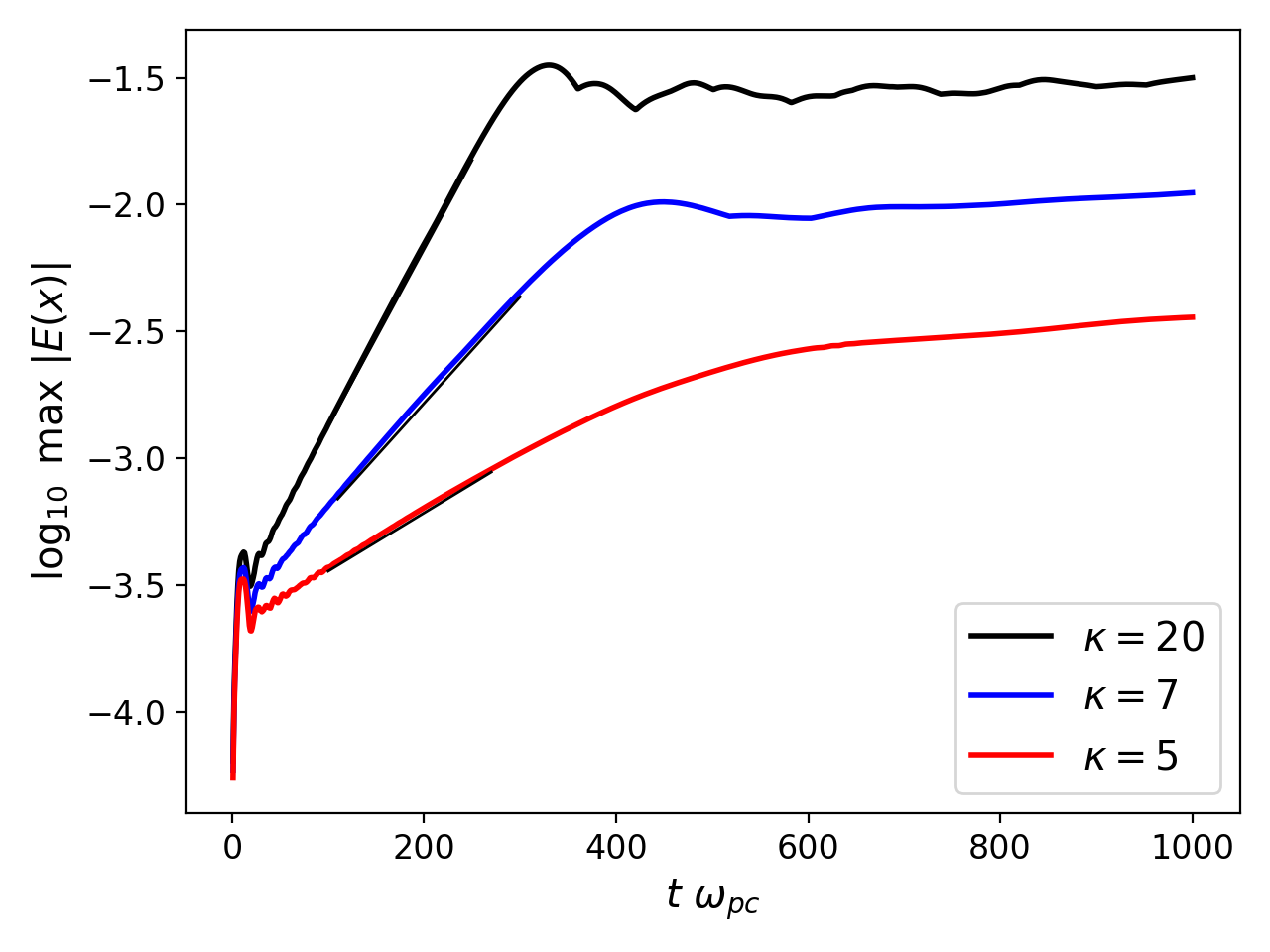}
\caption{Temporal evolution of the maximum electric field value for the three simulations with $\kappa$ distributed electrons. Derived growth rates (black lines) are $\gamma/ \omega_{pc}=0.0161, 0.0101,$ and $0.0053$ respectively. Simulation parameters are given in Table \ref{kappa_parameters}.} 
\label{electric}
\end{figure*}

We verify theoretical predictions with 1D1V Vlasov simulations, using the code described in \citet{Afify2024}. The box length is $L_x/ \lambda_{Dc}=50$, with grid spacing $\Delta x/ \lambda_{Dc} = 0.25$. Velocity spaces for proton core/beam and electrons are 12 times the respective thermal speeds, resolved with $151$ and $193$ points, respectively. Boundary conditions are periodic along $x$ and zero flux at $v= v_{min}, v_{max}$. We use the maximum value of the electric field in the simulation box as a measure for the growth of the instability. The same parameters as Figure \ref{DR_kappa}, panel a, are used for the simulation run depicted in Figure~\ref{electric}. The black lines superimposed to the linear phase of the instability mark the time interval used for growth rate calculations. A comparison between theoretical ($Th$) and simulated ($Sim$) growth rates at the simulated wavenumber $k \lambda_{Dc}= 2 \pi/ 50= 0.126$ is shown in Table \ref{kappa_parameters}, together with the real frequency of the instability and the resonance velocity given by Landau theory ($Th$).
The chosen simulated wavenumber is close to the maxima of the growth rate, see Fig. 2a.

\begin{figure*}  [h!]
     \centering
     \begin{minipage}[b]{0.8\textwidth}
    \includegraphics[width=\textwidth]{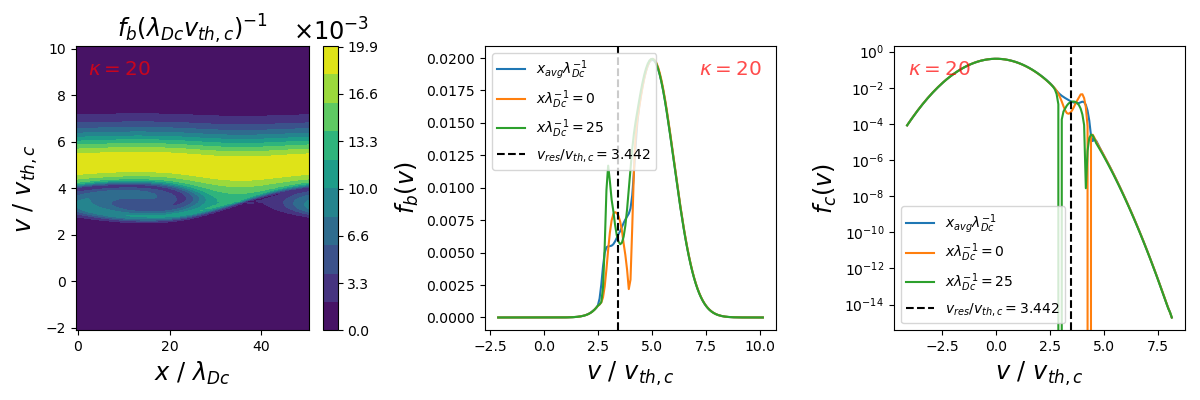}
  \end{minipage}
  \hfill
  \begin{minipage}[b]{0.8\textwidth}
    \includegraphics[width=\textwidth]{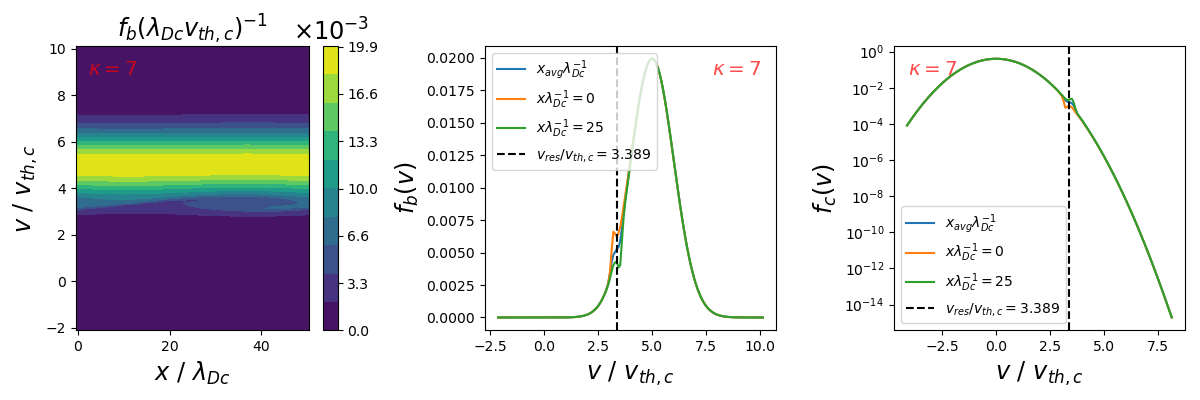}
  \end{minipage}
  \hfill
  \begin{minipage}[b]{0.8\textwidth}
    \includegraphics[width=\textwidth]{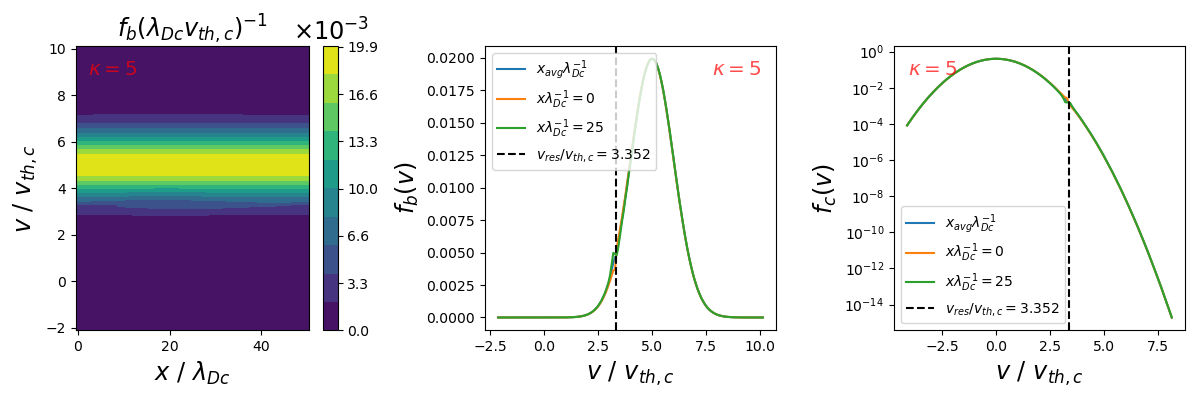}
  \end{minipage}
     \caption{Snapshots of the proton distributions at time $\omega_{pc}t= 300$, when all simulations are approaching the end of the linear phase, from simulations with electron $\kappa$ indices $\kappa = 20, \;7, \;5$. Left: beam phase-space distribution. Middle and right: beam and core velocity distributions, respectively, averaged in $x$ (blue lines), cut at $x/\lambda_{Dc}=0$ (orange line) and at $x/\lambda_{Dc}=25$ (green line). The vertical dashed lines indicate the theoretical estimates of the resonant velocity.} 
     \label{kappa-300}
\end{figure*}

\begin{figure*}  [h!]
     \centering
     \begin{minipage}[b]{0.8\textwidth}
    \includegraphics[width=\textwidth]{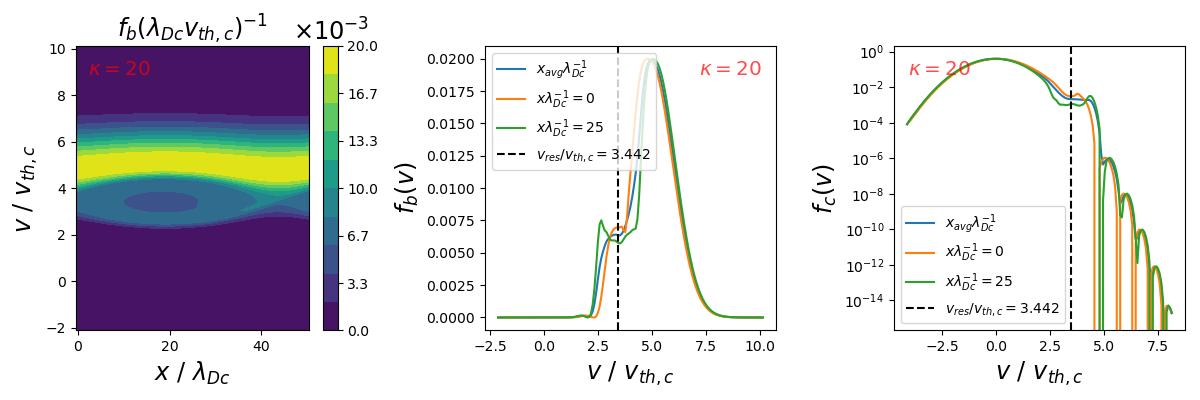}
  \end{minipage}
  \hfill
  \begin{minipage}[b]{0.8\textwidth}
    \includegraphics[width=\textwidth]{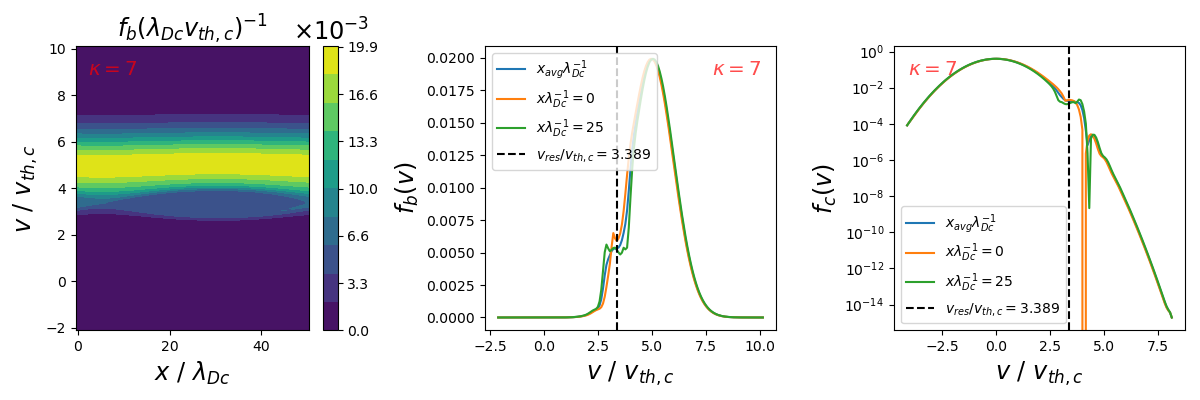}
  \end{minipage}
  \hfill
  \begin{minipage}[b]{0.8\textwidth}
    \includegraphics[width=\textwidth]{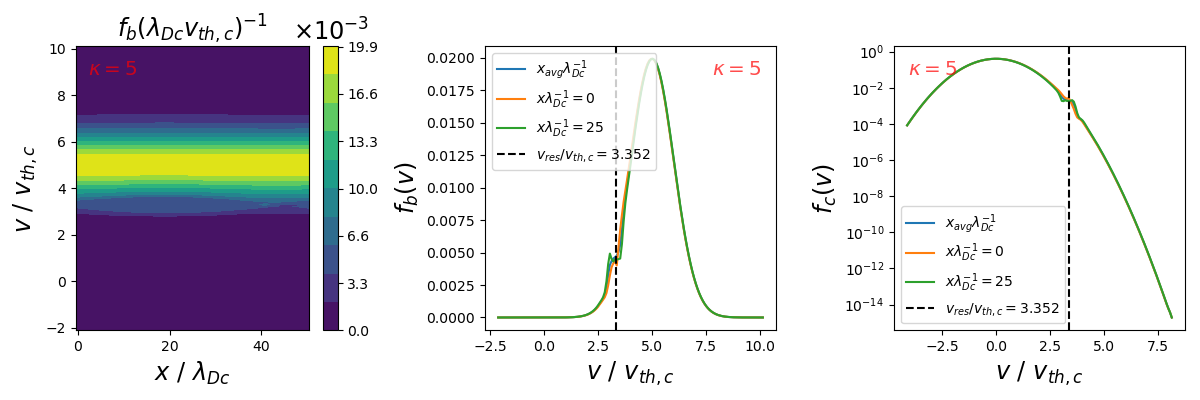}
  \end{minipage}
     \caption{As in Figure \ref{kappa-300}, at $ \omega_{pc}t = 1000$ when all instabilities have saturated.} 
     \label{kappa-1000}
\end{figure*}

Figure \ref{kappa-300} presents simulation results illustrating how $\kappa-$ distributed electrons modify the instability. These snapshots are taken at $\omega_{pc}t= 300$, when all simulations are approaching the end of the linear phase. 
The left columns depict the beam phase-space distribution function. We see traces of resonant beam interaction (ion hole formation), which are more developed at higher $\kappa$ values, where the instability has reached larger amplitude. 
In the middle and right columns we depict the beam and core VDFs, averaged in space, at $x/\lambda_{Dc}=0$ and at $x/\lambda_{Dc}=25$, respectively. The vertical dashed line indicates the resonance velocity, $v_{res}=\omega/k$, as calculated from linear theory at $k \lambda_{Dc}=0.126$. The averaged VD, as also the VD cuts at $x/ \lambda_{Dc}=0, \;25$, show that the VD is indeed modified in the vicinity of the resonant velocity. As the core protons are only very weakly affected by this resonant process, we plot their VD on a logarithmic scale. Figure \ref{kappa-1000} shows time $\omega_{pc}t= 1000$, when the instability in all simulations has already saturated. We observe the effect of resonance interaction on the proton beam population. We observe that at lower $\kappa$ values, signatures on both the proton core and beam population are weaker, consistent with smaller saturation amplitudes, see Fig.\ref{electric}

\section{Core-strahl electrons}
\label{sec:core-strahl}
We now consider an alternative distribution for the electron population, namely the core-strahl distribution often observed in the solar wind. In this Section, we use as temperature for the electron core ($ec$) population  $T_{ec}= 7 \;T_{c}$, as in~\citet{mozer2021triggered}. Notice that, with only one Maxwellian electron population at this temperature, the configuration is stable to the IIAI. We then redistribute part of these electrons into a hotter, Maxwellian strahl population, and examine the resulting effects on the IIAI. Notice that we now have four populations: two proton ($p$) and two electron ($e$) populations, each composed of a core ($c$) and a beam. In accordance with the nomenclature commen in solar wind physics, we label the proton beam with $b$, but the electron beam, usually called ''strahl", with $s$. In this notation, the densities are related by $n_{ec} + n_{es}= n_{b}+ n_{c}$. 
 We vary the density and temperature of the strahl population, and investigate the effects on the instability. In all these experiments, electron and proton cores are at rest. The density of the proton beam is kept fixed to the values analyzed in the previous Section, $n_{b}/ n_{c}=0.05$. The proton beam drift is kept at  $V_d/ v_{th,c}=5$, unless specified otherwise. The drift velocity of the strahl results from the zero current condition, $V_{d,es} =V_d n_{pb}/n_{es}$.

\begin{figure*}  [h!]
\begin{minipage}[b]{0.48\textwidth}
    \includegraphics[width=\textwidth]{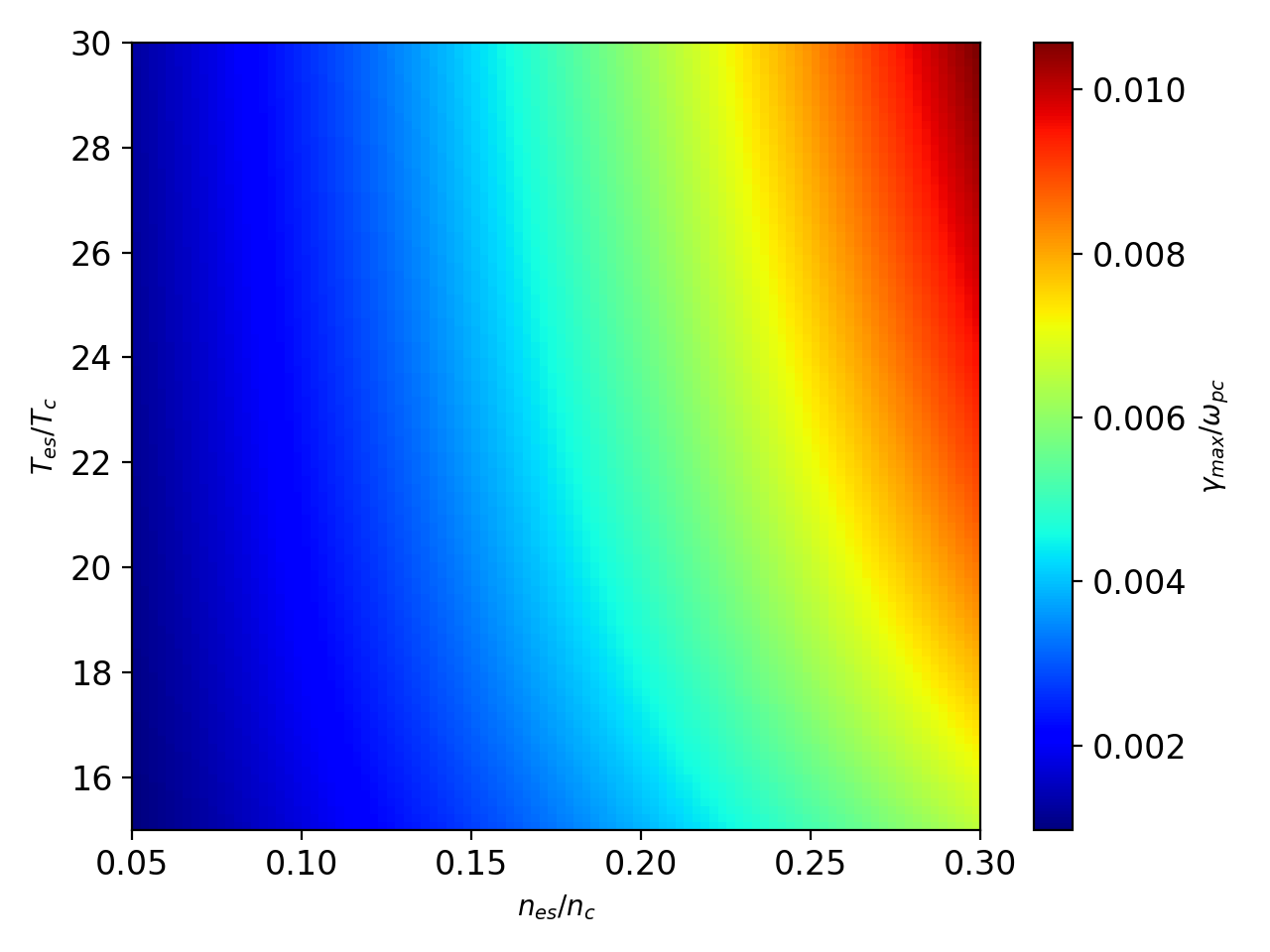}
  \end{minipage}
\begin{minipage}[b]{0.48\textwidth}
    \includegraphics[width=\textwidth]{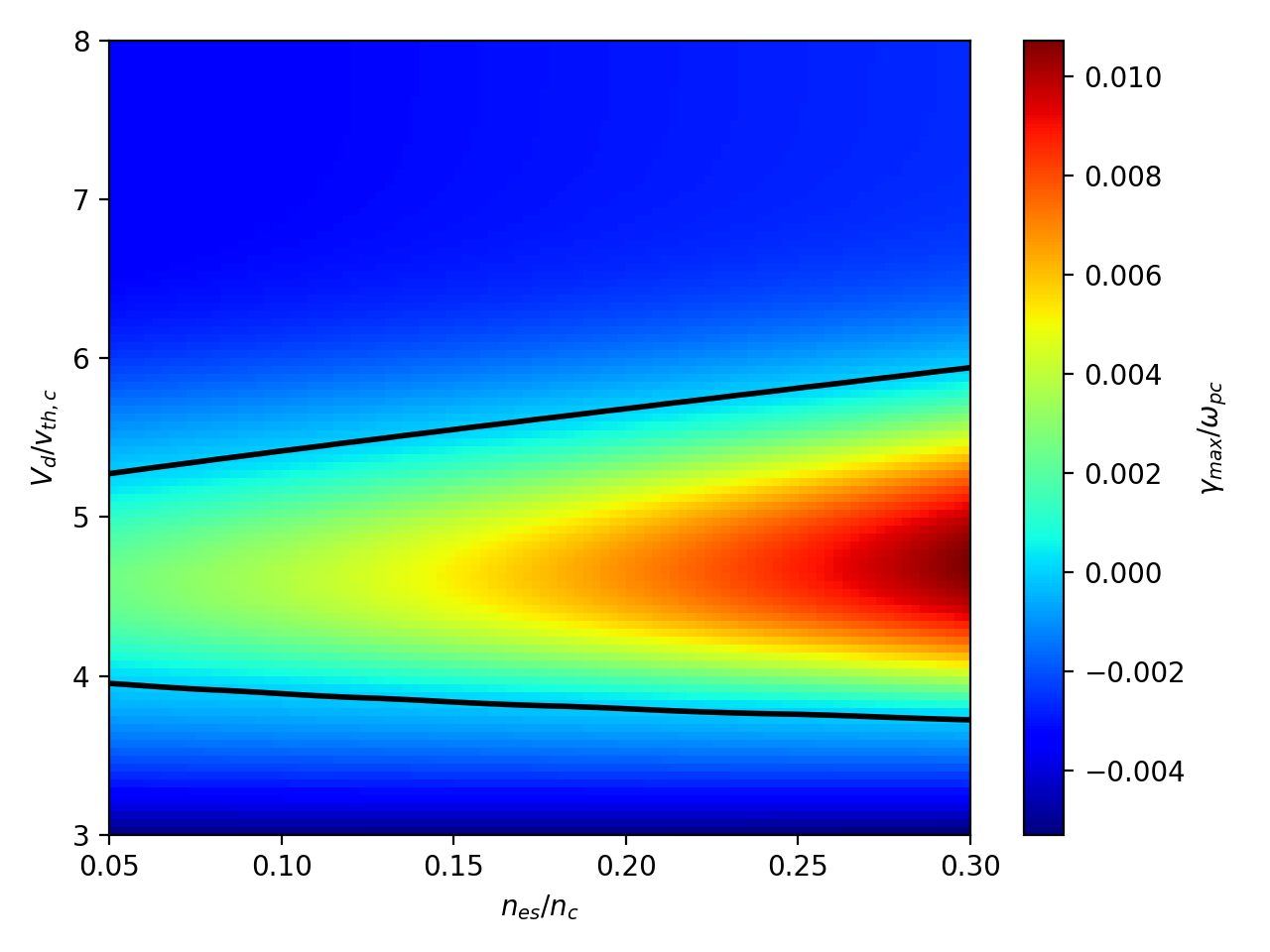}
  \end{minipage}
\caption{Maximum growth rate dependence on plasma parameters. 
Left panel: Maximum growth rate as a function of $n_{es}/n_{c}$ and $T_{es}/T_{c}$ for $V_{d}/v_{th,c}=5$, $n_{b}/n_{c}=0.05$, $T_{b}/T_{c}=1$, and $T_{ec}/T_{c}=7$. 
Right panel: Maximum growth rate as a function of $n_{es}/n_{c}$ and $V_{d}/v_{th,c}$ with the same parameters except $T_{es}/T_{c}=25$; solid lines indicate the $\gamma=0$ contour.}

\label{contour_coreStrahl}
\end{figure*}

Figure \ref{contour_coreStrahl} shows that a hotter strahl component tends to destabilize the instability, as shown by the maximum normalized growth rate, $\gamma_{\text{max}}/\omega_{pc}$, calculated from Landau theory and presented as a color-coded contour plot. The left panel illustrates the dependence of the growth rate on two key parameters: the electron strahl to core proton temperature ratio ($T_{es}/T_{c}$) and the electron strahl to core proton density ratio ($n_{es}/n_{c}$). A denser, hotter strahl favors instability onset. In the right panel we change only the strahl density, while keeping $T_{es}/ T_c= 25$. In the vertical axis, we allow the proton beam drift velocity, $V_{d}/v_{th,c}$, to change. As usual, see~\citet{Afify2024} and Fig.~\ref{DR_kappa} panel b, the instability only exists in a finite $V_{d}$ interval, which widens with increasing strahl density.
%Moreover, we notice that when $T_{es}/ T_c \rightarrow 7$ and $n_{es}\rightarrow 0$ (i.e., when the electrons are composed of a single population with $T_{es}/T_c=7$), the plasma is stable to ion acoustic instability, as we verified in~\citet{Afify2024}. Increasing strahl density and temperature increases the growth rate of the ensuing instability.
%\st{To identify the species responsible for driving IIAI, we present the phase arguments of each species in Fig. \ref{2electronsArgument}. The results indicate that the proton beam, characterized by a small $\zeta_{b} > 1$, serves as the primary driver. In contrast, both the strahl and core electrons exhibit stronger resonance at small $\zeta_{es,ec} < 1$, while the core protons remain less resonant due to their higher $\zeta_{c} > 1$.}

\begin{figure*}  [h!]
\centering
\begin{minipage}[b]{0.45\textwidth}
    \includegraphics[width=\textwidth]{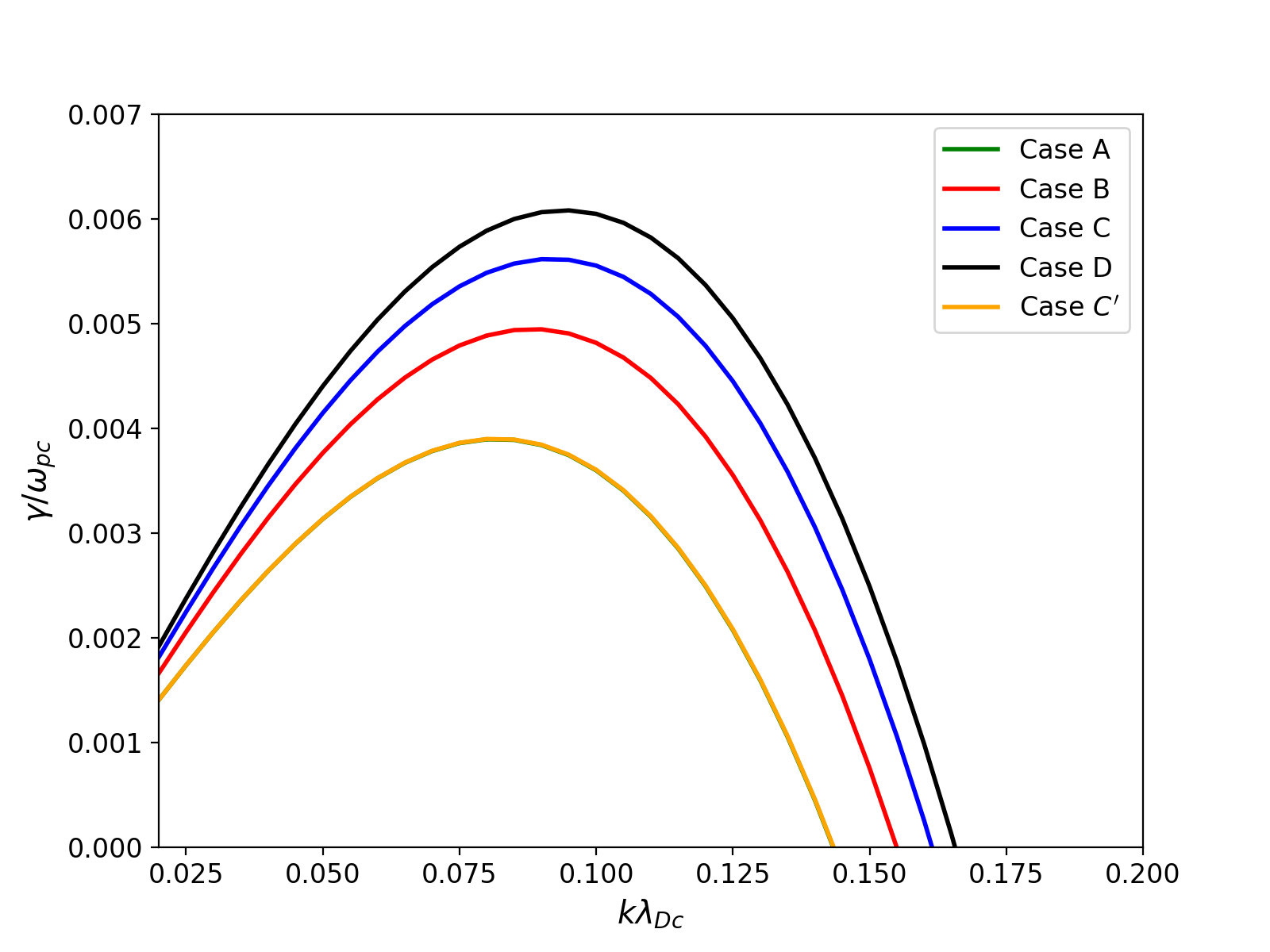}
  \end{minipage}
\begin{minipage}[b]{0.45\textwidth}
\includegraphics[width=\textwidth]{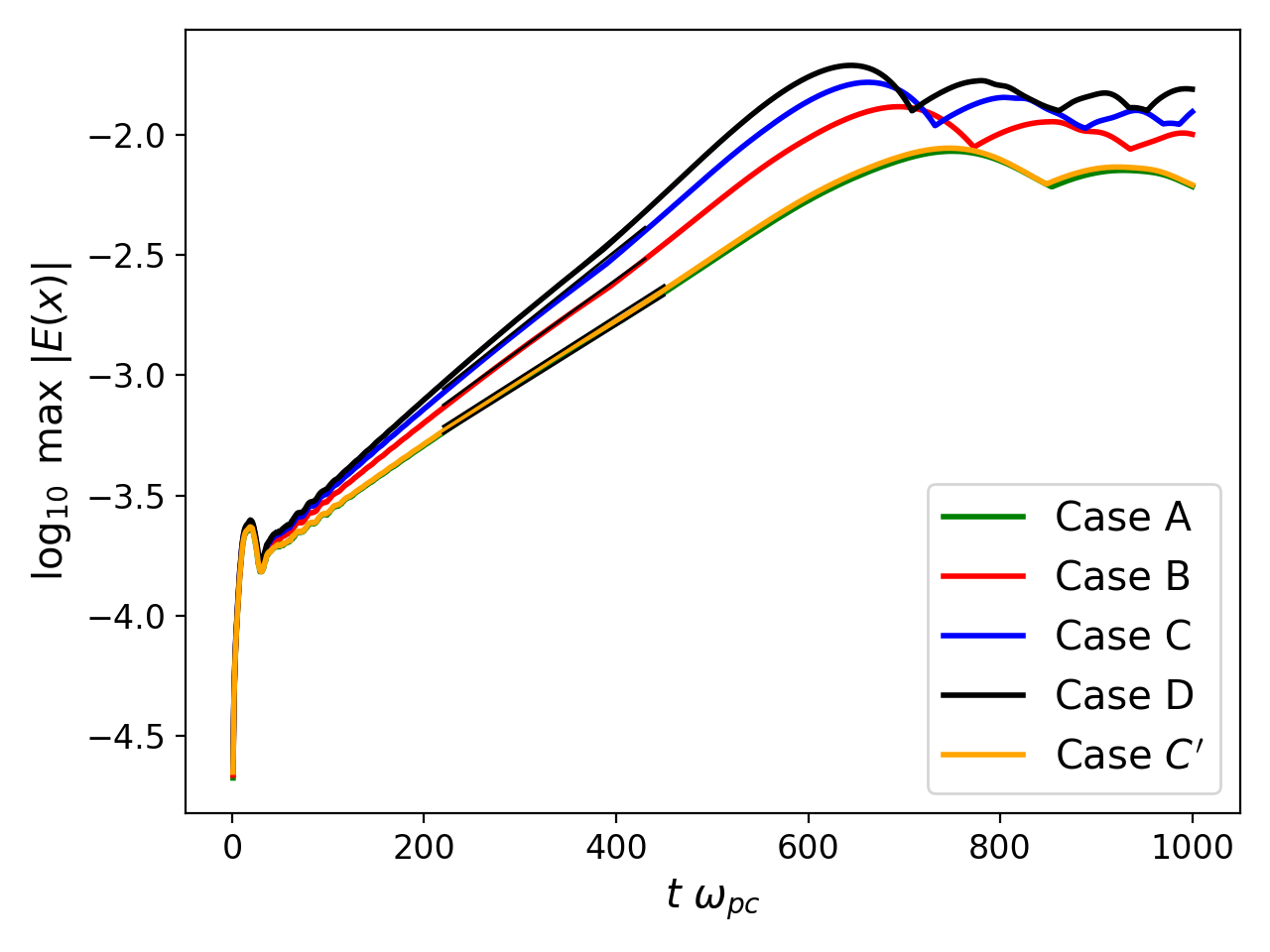}
\end{minipage}
\caption{Dispersion and electric field evolution for the IIAI. 
Left panel: Dispersion relation for the IIAI with parameters given in Table \ref{tab:SC_runs}. Right panel: Simulated electric field evolution for the same cases. Common parameters to all cases are $n_{b}/n_{c}=0.05$, $T_{b}/T_{c}=1$, $V_{d}/v_{th,c}=5$, $T_{ec}/T_{c}=7$, with electron and proton cores at rest.}
 \label{SC_Sim}
\end{figure*}

In Table~\ref{tab:SC_runs} we list a number of cases, where we change the strahl temperature with fixed strahl density $n_{es}/ n_{c}=0.2$ (cases A-D), and one ($\mathrm{Case}\ C^\prime$) when we reduce strahl density with respect to case C.
As before, we provide real frequency and growth rate calculated at $k \lambda_{Dc}= 0.08$, from linear theory, and the growth rate obtained from simulations with box size $L_x/ \lambda_{Dc}= 80$. 
Similarly, the chosen simulated wavenumber corresponds to the maximum growth rates as inferred from linear theory.
In Figure \ref{SC_Sim}, we provide the full dispersion relation from theory, panel a, and the time history of the maximum electric field values from simulations, panel b, for the cases in Table~\ref{tab:SC_runs}.

\begin{figure*}  [h!]
\centering
\begin{minipage}[b]{0.5\textwidth}
    \includegraphics[width=\textwidth]{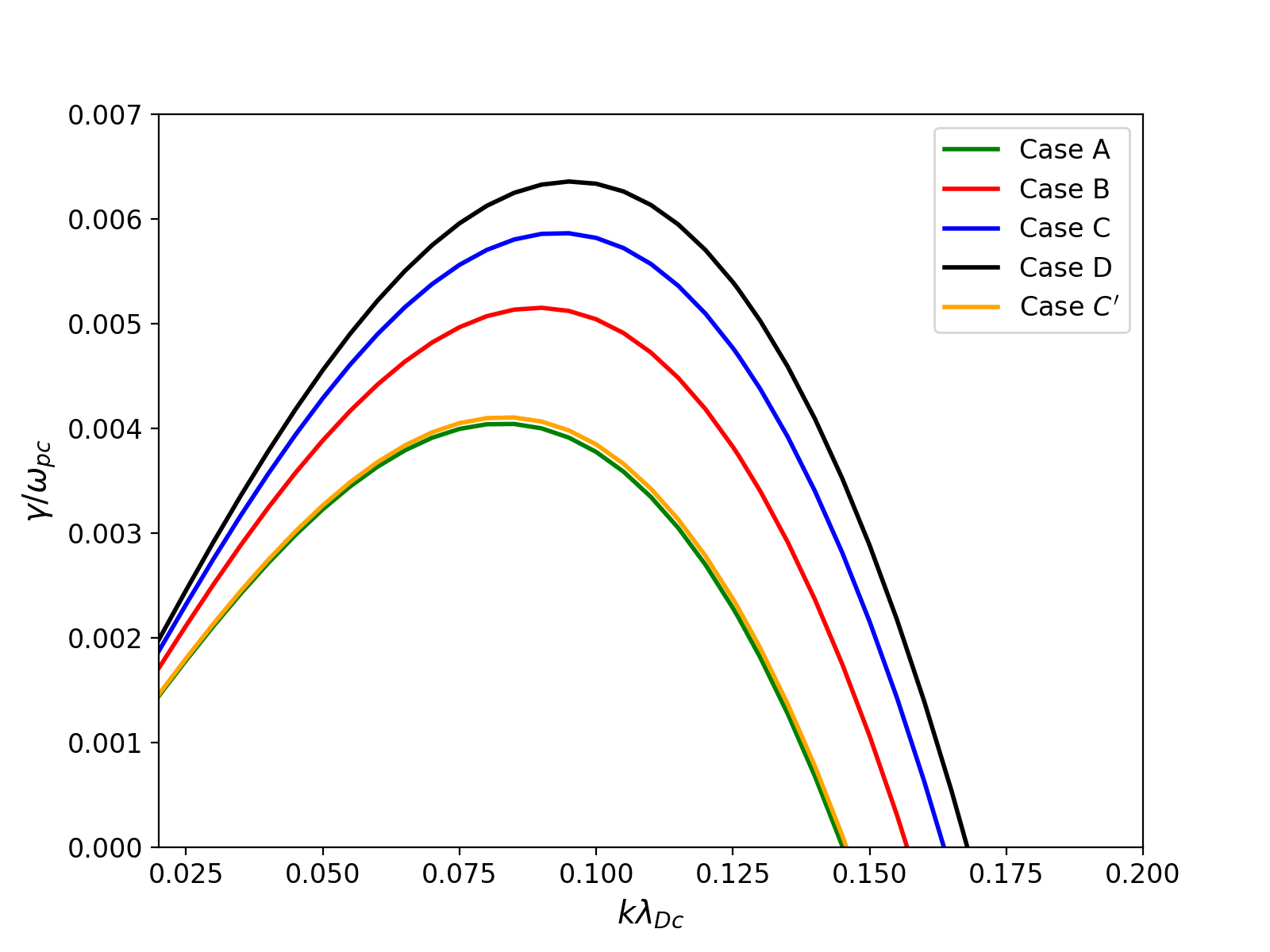}
  \end{minipage}
\begin{minipage}[b]{0.48\textwidth}
\includegraphics[width=\textwidth]{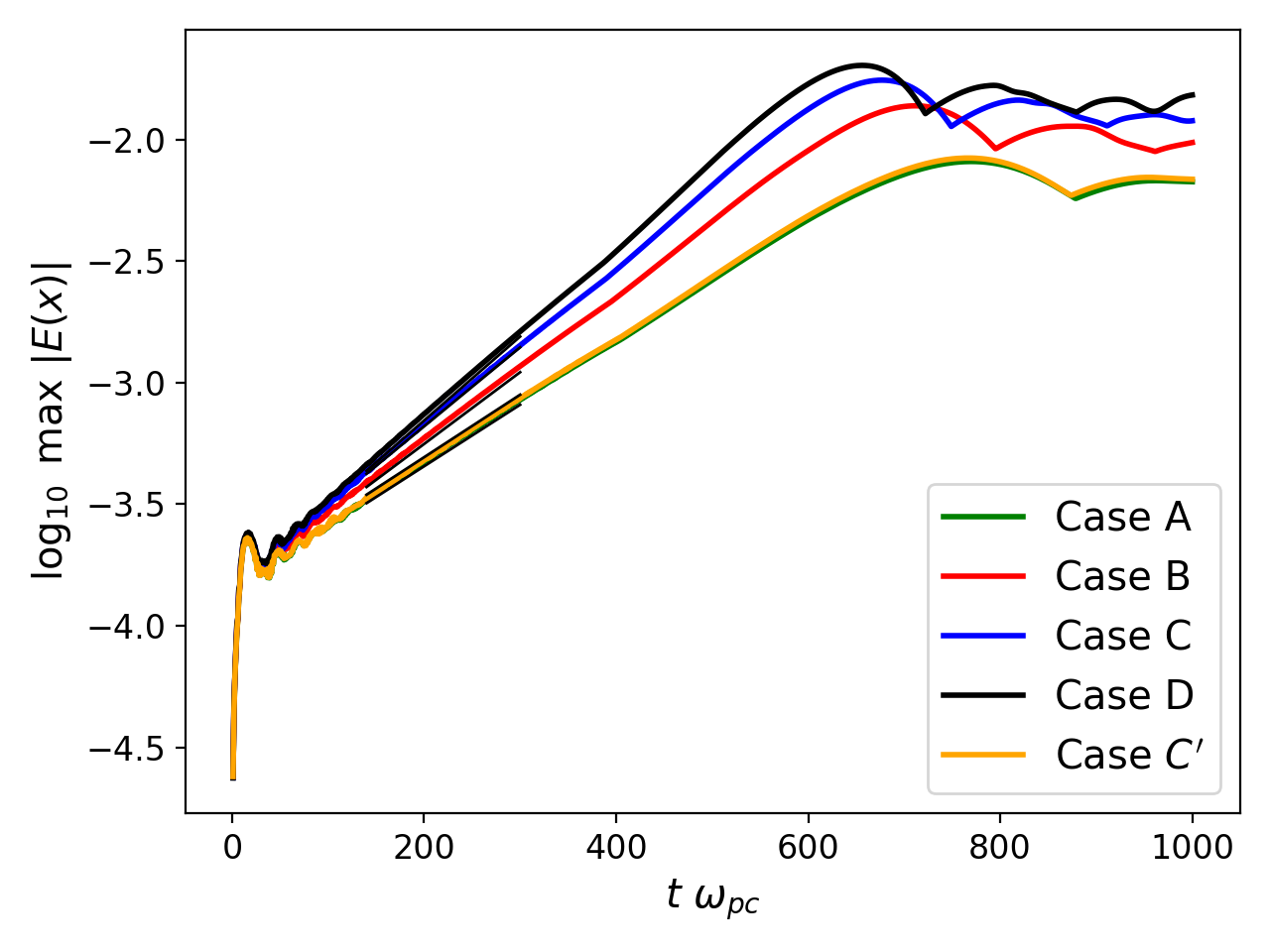}
\end{minipage}
\caption{As in Fig.~\ref{SC_Sim}, but with a single, Maxwellian electron distribution with $T_e= T_{eff}$.} 
\label{SC_eff}
\end{figure*}

\begin{figure*}  [h!]
     \centering
     \begin{minipage}[b]{0.7\textwidth}
    \includegraphics[width=\textwidth]{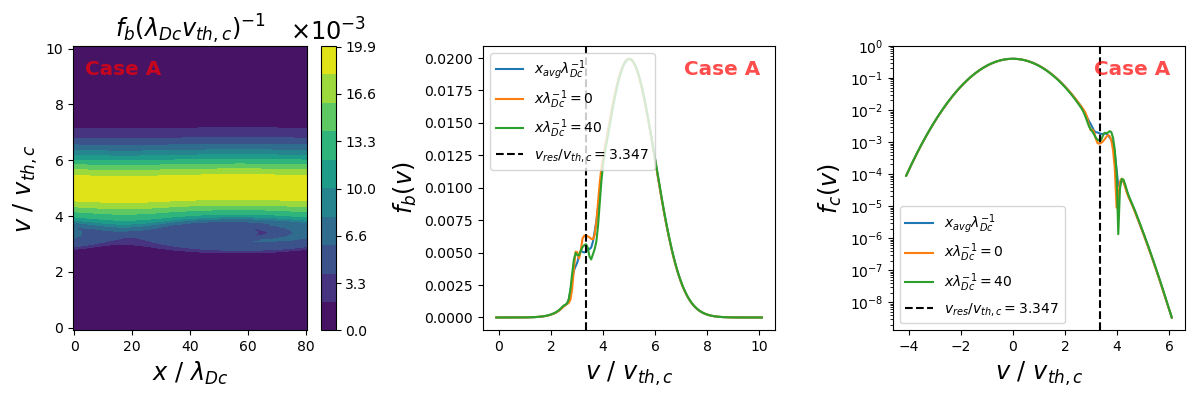}
  \end{minipage}
  \hfill
  \begin{minipage}[b]{0.7\textwidth}
    \includegraphics[width=\textwidth]{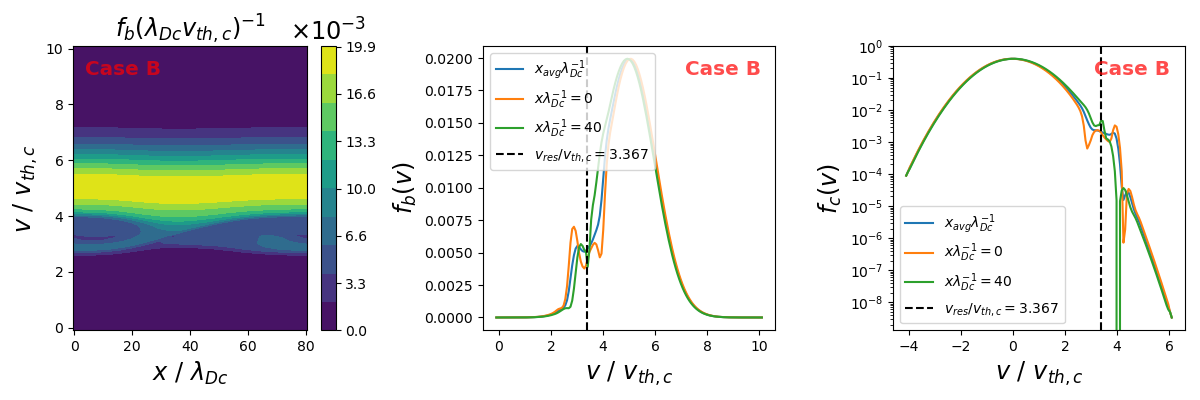}
  \end{minipage}
  \hfill
  \begin{minipage}[b]{0.7\textwidth}
    \includegraphics[width=\textwidth]{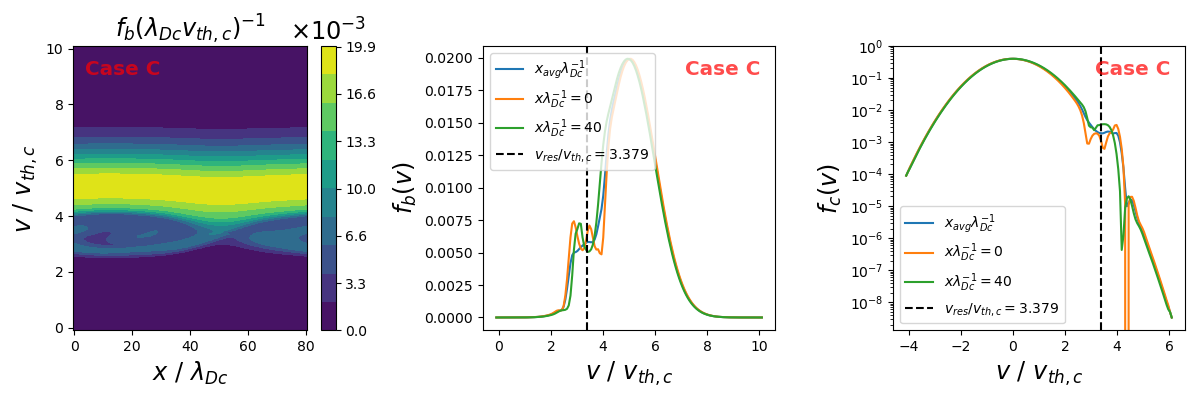}
  \end{minipage}
  \hfill
  \begin{minipage}[b]{0.7\textwidth}
    \includegraphics[width=\textwidth]{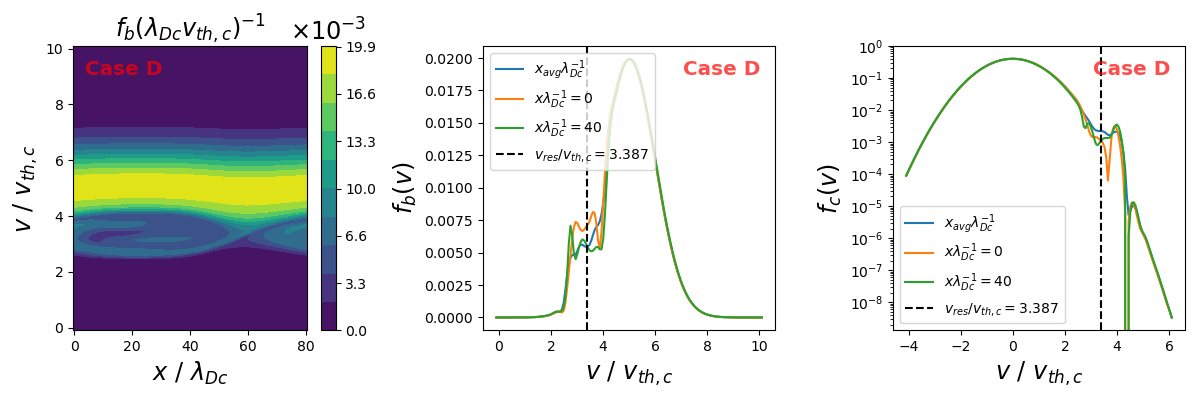}
  \end{minipage}
  \hfill
  \begin{minipage}[b]{0.7\textwidth}
    \includegraphics[width=\textwidth]{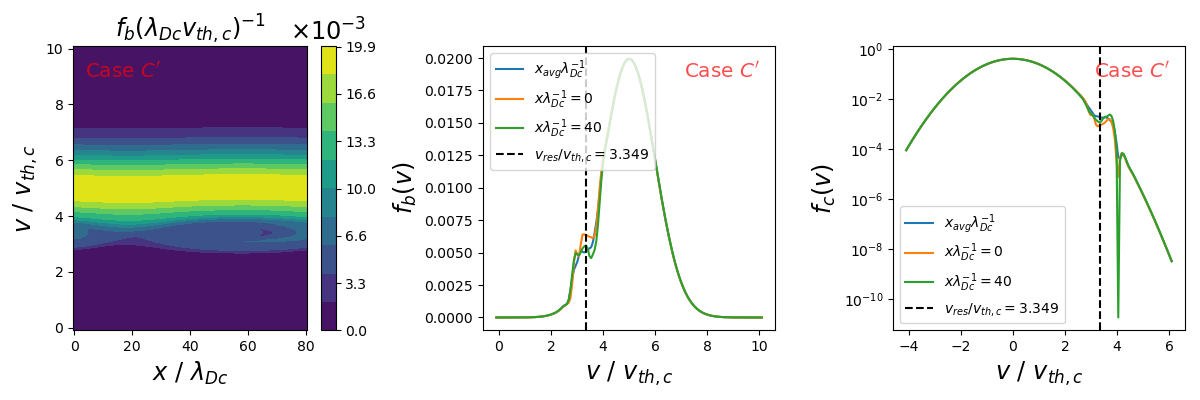}
  \end{minipage}
     \caption{Snapshots of the core and beam distribution as a function of the strahl's temperature for the cases listed in Table \ref{tab:SC_runs} at time $\omega_{pc}t= 1000$, when all simulations are saturated. Left: beam phase-space distribution. Middle and right: beam and core velocity distributions, respectively, averaged in $x$ (blue lines), cut at $x/\lambda_{Dc}=0$ (orange line) and at $x/\lambda_{Dc}=40$ (green line). The vertical dashed lines indicate the theoretical estimates of the resonant velocity.} 
     \label{strahlPhase-1000}
\end{figure*}

\begin{figure*}  [h!]
     \centering
     \begin{minipage}[b]{0.7\textwidth}
    \includegraphics[width=\textwidth]{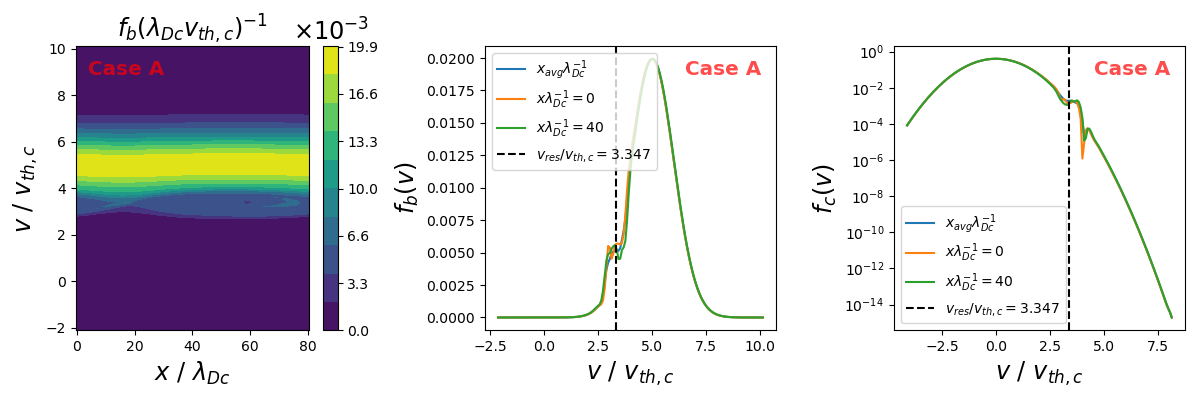}
  \end{minipage}
  \hfill
  \begin{minipage}[b]{0.7\textwidth}
    \includegraphics[width=\textwidth]{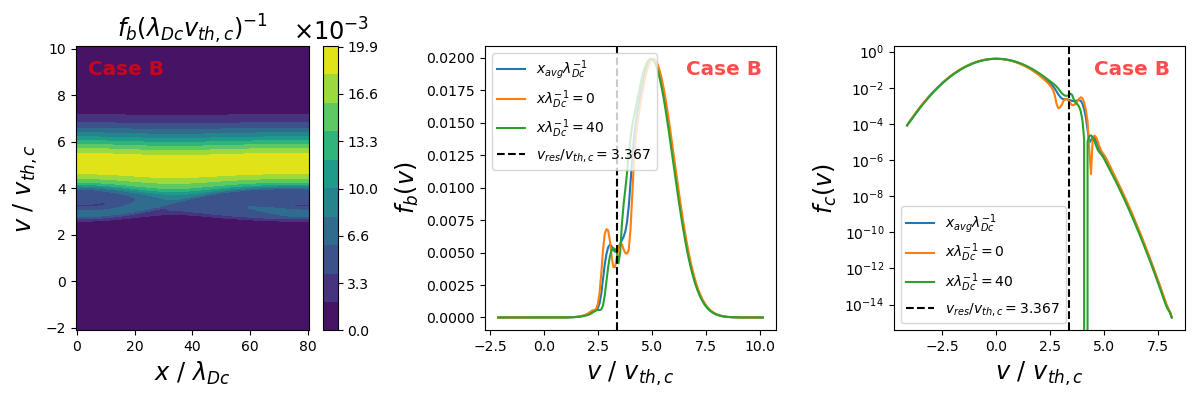}
  \end{minipage}
  \hfill
  \begin{minipage}[b]{0.7\textwidth}
    \includegraphics[width=\textwidth]{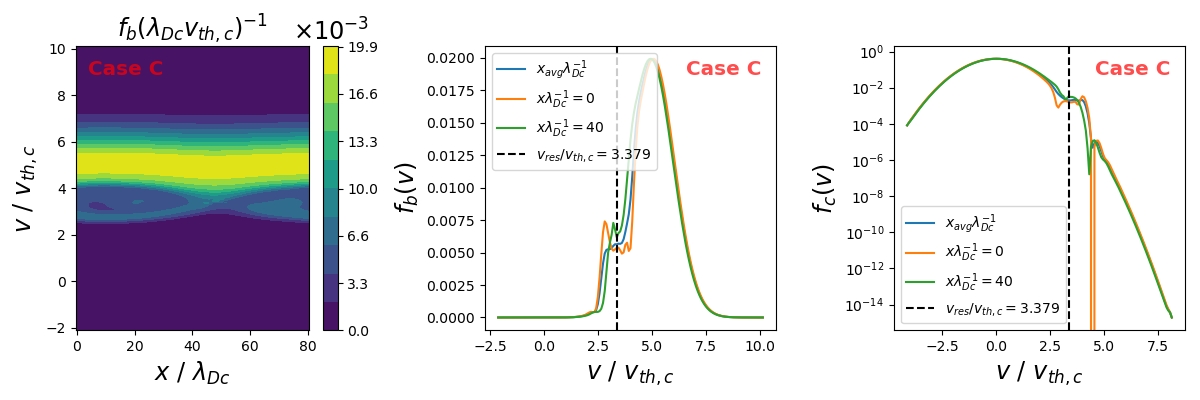}
  \end{minipage}
  \hfill
  \begin{minipage}[b]{0.7\textwidth}
    \includegraphics[width=\textwidth]{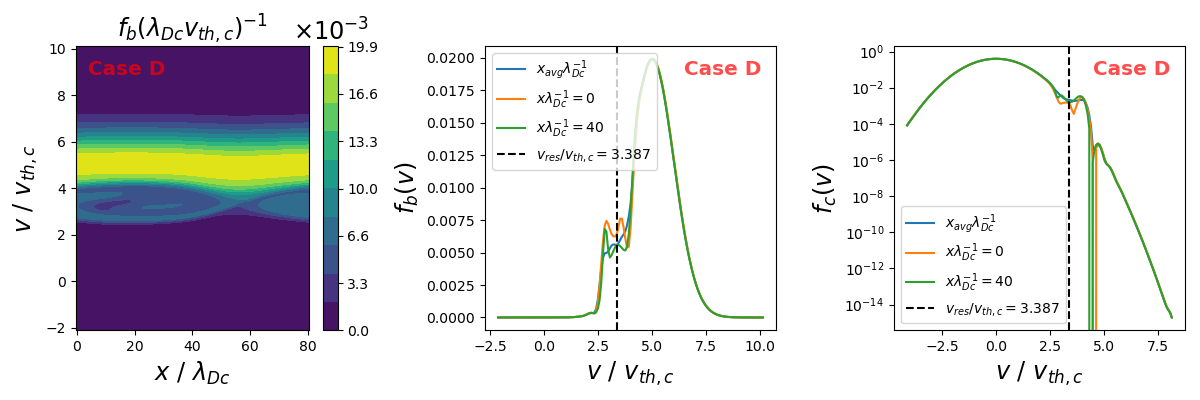}
  \end{minipage}
  \hfill
  \begin{minipage}[b]{0.7\textwidth}
    \includegraphics[width=\textwidth]{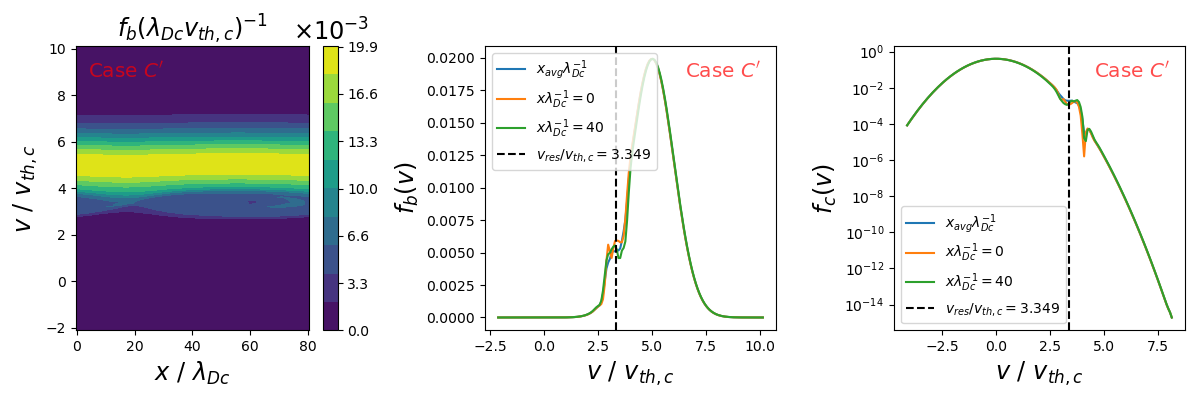}
  \end{minipage}
     \caption{As in Fig.~\ref{strahlPhase-1000}, but with a single, Maxwellian electron distribution with $T_e= T_{eff}$.}
     %\caption{Evolution of the core and beam velocity distribution function as a function of the core electron's effective temperature. All plots are at time $\omega_{pc}t= 1000$, where all simulations pass the linear phase.  } 
     \label{effective-1000}
\end{figure*}

%In the Series 1 runs, we increase the strahl temperature, keeping its density fixed. In the Series 2 runs, we increase the strahl density, keeping its temperature fix.
We observe a good agreement between calculated and simulated growth rate, with a tendency of faster growths in simulations with respect to theory.
~\citet{jones1975propagation}  calculates an effective electron temperature, $T_{eff}$, for IIAI evolution for an electron distribution composed of core and strahl: 
\begin{equation}
T_{eff}= (n_{e} T_{es} T_{ec})/(n_{es} T_{ec}+ n_{ec}T_{es}).
\label{eq:Gary_eff}
\end{equation} 
We calculate $T_{eff}$ for our reference cases in Table~\ref{tab:SC_runs}. 
In Table~\ref{tab:Effec_Tem} and Fig.~\ref{SC_eff}, we repeat the dispersion relation calculations and simulations as in Table~\ref{tab:SC_runs} and Fig.~\ref{SC_Sim}, using as electron distribution a single Maxwellian with the temperature $T_{eff}$ calculated from the corresponding core-strahl cases. We verify that theoretical results are identical and simulated results are fairly close, thus validating the concept of effective temperature.
This is underpinned by the comparison of proton beam and core distributions in Fig. \ref{strahlPhase-1000} (core-strahl electron distribution) and Fig.\ref{effective-1000} (Maxwellian electron distribution, with $T_e= T_{eff}$). We observe the same wave-particle interaction patterns in both cases.

\section{Discussion and conclusions}
\label{sec:concl}

This study examines the influence of nonthermal electron populations, $\kappa$ and core-strahl, on the triggering and evolution of the IIAI. 

This study was motivated by our previous results,~\citet{Afify2024}, where we found that purely Maxwellian electron distribution would need slightly larger temperatures than what observed in~\citet{mozer2021triggered} to allow for the observed IIAI-related wave activity. 

The electron velocity distribution functions that we considered here are $\kappa$ and core-strahl, which are customarily used to model observed electron VDFs in the solar wind. 
We find that $\kappa$-distributed electron tend to stabilize the IIAI, with smaller growth rates at lower $\kappa$ indices. Adding a strahl distribution acts in the direction of destabilizing the IIAI with respect to a Maxwellian distribution. Stronger instabilities are observed for hotter and denser strahl populations. 

Addressing the relationship with observations, we note that the strahl temperature of $T_{es}/T_c\approx 15-20$ used in our study corresponds to $T_{es}$ being $2$ to $3$ times the electron core temperature and thereby might be fairly realistic for the solar wind. However, our $n_{es}/n_c = 0.15 - 0.2$ assumes a much denser strahl than the values of $0.05 - 0.1$ reported by \citet{maksimovic2005radial} for solar distances of $0.3-1$ AU. 
Therefore, this study does not explain why IIAI-related wave activity is observed in~\citet{mozer2021triggered}.
On the other hand, \citet{maksimovic2005radial} also reported the strahl density to decrease with increasing solar distance, so that a more massive electron strahl at PSP positions around 20 $R_{\odot}$ might be possible.

Physical insights that we obtain from this work are the following. Lower $\kappa$ values tend to stabilize the IIAI due to the higher electron phase-space density at $v \sim v_{res} << v_{th, ec}$ (see inset of Fig.~\ref{VDF_kappa}, panel a), leading to enhanced Landau damping in the electrons. Since our parameters are very close to the instability threshold, even a slightly enhanced electron damping can determine instability onset.
In the case of the core-strahl distribution, redistributing core electrons into a hotter strahl population reduces Landau damping, and slightly enhances the growth rate. This effect is captured in the formulation for an effective temperature provided by~\citet{jones1975propagation}, and validated here with simulations. Using this concept, and taking the limit of infinitivley hot strahl electrons, with a realistic strahl density $n_{es}= 0.05 n_e$, the effective electron temperature would increase by only a few percents with respect to the electron core temperature, see Eq.~\ref{eq:Gary_eff} and discussion in~\citet{jones1975propagation}. We then speculate that the IIAI activity observed in~\citet{mozer2021triggered} may be related to external drivers not captured in our model, for example, temporary beam density or drift enhancements e.g. due to reconnection events, in conjunction with increasing electron to proton temperature ratio, rather than to the specifics of the electron velocity distribution.

\section*{Acknowledgement}
M. S. Afify thanks the Alexander-von-Humboldt Foundation, 53173 Bonn, Germany (Ref 3.4-1229224-EGY-HFST-P) for the research fellowship and its financial support. M.E.I. acknowledges support from the Deutsche Forschungsgemeinschaft (DFG, German Research Foundation) within the Collaborative Research Center SFB1491 and project 497938371. We thank J.Verniero and J.Halekas for useful discussion regarding PSP observations.

\begin{table*}
\caption{Theoretical ($Th$) growth rates, frequencies, and resonant velocities $v_{res}=\omega/k$, and observed simulation growth rates ($Sim$) for $\kappa$ distributed electrons and proton beam drift speed  $V_{d}/ v_{th,c}= 5$, proton beam density $n_b/n_c=0.05$, beam temperature $T_b/ T_c= 1.0$, electron temperature $T_e/T_c=10$, and wavenumber $k\lambda_{Dc}=0.126$. }
\label{kappa_parameters}
\centering
\begin{tabular}{c c c c c c c}
\hline\hline  
         Name     & $\kappa$ & $(\omega/\omega_{pc})_{Th}$ & $(v_{res}/ v_{th,c})_{Th}$ & $(\gamma/\omega_{pc})_{Th}$ & $(\gamma/\omega_{pc})_{Sim}$\\
    \hline
        Case 1 &  20 & 0.434 & 3.442 & 0.0148 & 0.0161\\ 
        Case 2 & 7 & 0.427 & 3.389 & 0.0101 & 0.0101\\ 
        Case 3 &    5 & 0.422 & 3.352 & 0.0068 & 0.0053\\ 
    
    \hline
\end{tabular}
\end{table*}

\begin{table*}
\caption{Theoretical ($Th$) growth rates, frequencies, and resonant velocities $v_{res}=\omega/k$, and observed simulation growth rates ($Sim$) for core-strahl distributed electrons for various strahl densities and temperatures. Parameters kept fixed across all cases are  $V_{d}/v_{th,c}=5$, $n_{b}/n_{c}=0.05$, $T_b/T_c=1.0$,  and $k\lambda_{Dc}=0.08$. }
\label{tab:SC_runs}
\centering
\begin{tabular}{c c c c c c c c}
    \hline\hline  
         Name & $\frac{n_{es}}{n_{c}}$  & $\frac{T_{es}}{T_c}$   &  $(\omega/\omega_{pc})_{Th}$ & $(v_{res}/ v_{th,c})_{Th}$ &$(\gamma/\omega_{pc})_{Th}$ & $(\gamma/\omega_{pc})_{Sim}$\\
    \hline
        Case A & 0.2  & 15.0 &   0.268 & 3.347 & 0.0039 & 0.00576\\ 
        Case B & 0.2  & 20.0 &   0.270 & 3.367 & 0.0049 & 0.00668\\ 
        Case C & 0.2  & 25.0 &   0.270 & 3.379 & 0.0055 & 0.00737\\ 
        Case D & 0.2  & 30.0 &   0.271 & 3.387 & 0.0059 & 0.00783\\ 
        %Case 2a & 0.1  & 25.0 & 9.03 & -0.000327 & 0.0023\\ 
        $\mathrm{Case}\ C^\prime$ & 0.15  & 25.0  & 0.268 & 3.349 & 0.0039 & 0.00587\\ 
    
    \hline
\end{tabular}
\end{table*}

\begin{table*}
\caption{As in Table~\ref{tab:SC_runs}, using a single Maxwellian electron distribution with $T_e=T_{eff}$ instead of a core-strahl distribution.}
\label{tab:Effec_Tem}
    \centering
    \begin{tabular}{c c c c c c c}
    \hline\hline
         Name &  $\frac{T_{eff}}{T_c}$ & $(\omega/\omega_{pc})_{Th}$ & $(v_{res}/ v_{th,c})_{Th}$ &$(\gamma/\omega_{pc})_{Th}$ & $(\gamma/\omega_{pc})_{Sim}$\\
    \hline
        Case A &   7.792 & 0.268 & 3.347 & 0.004 & 0.00583\\ 
        Case B &   7.989 & 0.269 & 3.367 & 0.0051 & 0.00679\\ 
        Case C &   8.113 & 0.270 & 3.379 & 0.0057 & 0.00748\\ 
        Case D &  8.197 & 0.271 & 3.387 & 0.0061 & 0.00794\\ 
        $\mathrm{Case}\ C^\prime$ &   7.803 & 0.268 & 3.349 & 0.0041 & 0.00592\\ 
    
    \hline
    \end{tabular}
\end{table*}

%\begin{figure}[h!]
%\centering
%\begin{minipage}[b]{0.4\textwidth}
   % \centering
    %\includegraphics[width=\textwidth]{beam_kappaArgument.png}
%\end{minipage}
%\vspace{1em} % vertical spacing between rows
%\begin{minipage}[b]{0.4\textwidth}
%    \centering
 %   \includegraphics[width=\textwidth]{core_kappaArgument.png}
%\end{minipage}
%\hfill
%\begin{minipage}[b]{0.4\textwidth}
 %   \centering
  %  \includegraphics[width=\textwidth]{electron_kappaArgument.png}
%\end{minipage}
%\vspace{0.5em}
%\caption{The resonant arguments as a function of the normalized wavenumber ($k\lambda_{D,c}$), based on the parameters specified in Fig.~\ref{VDF_kappa}. \textcolor{red}{REMOVE THIS ONE}}
%\label{kappaArguments}
%\end{figure}

%\begin{figure*}  [h!]
%\centering
%  \begin{minipage}[b]{0.45\textwidth}
%\includegraphics[width=\textwidth]{2electrons_arguments.png}
%\end{minipage}
%\caption{Argument resonance as a function of the normalized wavenumber, for $T_{eb}/T_c=20$, $T_{ec}/T_c=7$, $n_{eb}/n_c=0.2$, $V_ec/V_{th,c}=0$,  $V_c/V_{th,c}=0$, $V_b/V_{th,c}=5$, and $n_{b}/n_c=0.05$. \textcolor{red}%{REMOVE THIS}} 
%\label{2electronsArgument}
%\end{figure*}

\clearpage
\section*{References}
\bibliographystyle{aa}
\bibliography{references2}
\end{document}